\documentclass[aps,twocolumn,showlabels,showrefs,amsmath,amssymb,prl,superscriptaddress,floatfix,colors]{revtex4-1}

\usepackage{graphicx}
\usepackage{dcolumn}
\usepackage{bm}
\usepackage{amssymb}
\usepackage{hyperref}
\usepackage{multirow}
\usepackage{color}
\usepackage{ulem}
\usepackage{scalerel}
\usepackage{scalerel}
\usepackage[usestackEOL]{stackengine}

\def\Xint#1{\mathchoice
    {\XXint\displaystyle\textstyle{#1}}%
    {\XXint\textstyle\scriptstyle{#1}}%
    {\XXint\scriptstyle\scriptscriptstyle{#1}}%
    {\XXint\scriptscriptstyle\scriptscriptstyle{#1}}%
      \!\int}
\def\XXint#1#2#3{{\setbox0=\hbox{$#1{#2#3}{\int}$}
    \vcenter{\hbox{$#2#3$}}\kern-.5\wd0}}
\def\dashint{\Xint-}

\begin{document}

\title{(Non equilibrium) Thermodynamics of Integrable models: \\
The Generalized Gibbs Ensemble description of the classical Neumann Model
}

\date{\today}

\author{Damien Barbier}
\affiliation{Sorbonne Universit\'e, Laboratoire de Physique Th\'eorique et Hautes Energies, CNRS UMR 7589,
4 Place Jussieu, 75252 Paris Cedex 05, France}

\author{Leticia F. Cugliandolo}
\affiliation{Sorbonne Universit\'e, Laboratoire de Physique Th\'eorique et Hautes Energies, CNRS UMR 7589,
4 Place Jussieu, 75252 Paris Cedex 05, France}
\affiliation{Institut Universitaire de France, 1 rue Descartes, 75005 Paris France}

\author{Gustavo S. Lozano}
\affiliation{Departamento de F\'{\i}sica, Universidad de Buenos Aires, and  IFIBA CONICET, Argentina}

\author{Nicol\'as Nessi}
\affiliation{Departamento de F\'{\i}sica, Universidad Nacional de La Plata, and
IFLP CONICET, Diag 113 y 64 (1900) La Plata,
Argentina}


\begin{abstract}
 We study a classical integrable (Neumann) model describing the motion of a particle on the $S_{N-1}$ sphere, 
 subject to harmonic forces.
 We tackle the problem in the $N\rightarrow\infty$ limit by introducing a soft version in which the spherical constraint is imposed only on average over  initial conditions. We show that the Generalized Gibbs Ensemble 
 captures the long-time averages of the soft model. 
 We reveal the full dynamic phase diagram with extended, quasi-condensed,  coordinate-, and coordinate and momentum- condensed phases. The scaling properties of the fluctuations
allow us to establish in which cases the strict and soft spherical constraints are equivalent, confirming the validity of the GGE hypothesis for the Neumann model on a large portion of the dynamic phase diagram.
\end{abstract}

\maketitle

Interest in the long time dynamics of quantum isolated systems has continuously grown since
the celebrated quantum Newton's cradle experiment~\cite{Kinoshita06},
which proved that a quenched one-dimensional Bose gas does not reach standard thermal equilibrium.
Soon after, a Generalized Gibbs Ensemble (GGE) was proposed  to
describe typical observables in the steady state of systems with an extensive number of conserved quantities, say
$I_\mu$ with $\mu=1,\dots, N$~\cite{Rigol07,Rigol08}.
The pertinence of such density matrix was studied in a myriad of different
cases~\cite{Polkovnikov10,Pasquale-ed16,Gogolin16,Prosen15}.

Although most studies of quenches of isolated systems have focused on quantum systems,
non-ergodic dynamics are not  specifically quantum:
classical integrable systems~\cite{Khinchin,Dunajski12,Arnold78} are not expected to
reach equilibrium as dictated by conventional statistical mechanics either. One
can then ask whether a GGE description could apply to their long-term evolution as well
and, if so, under which conditions.
Yuzbashyan argued that the Generalized Microcanonical  Ensemble (GME),
in which the value of all constants of motion are fixed, is exact for classical integrable systems~\cite{Yuzbashyan16}.
However, this does not ensure that a canonical GGE  could be derived from the GME,
especially in long-range interacting systems for which the additivity of the conserved quantities is not
justified~\cite{Campa09,Dauxois10}. It is therefore of paramount importance to explicitly construct
the GGE of a classical integrable interacting model and put to the test its main statement,
that in the stationary limit~\cite{footnote} the long-time average,
$\overline{A} = \lim_{\tau\to\infty} \lim_{t_{\rm st} \gg t_0}\tau^{-1} \int_{t_{\rm st}}^{t_{\rm st} + \tau} \! dt' \, A(t')$, and
the phase space average,
$\langle A \rangle_{\scaleto{\rm GGE}{4pt}}  =$
$   \sum_{\rm conf} A e^{-\sum_\mu \gamma_\mu I_\mu({\rm conf})}/Z_{\scaleto{\rm GGE}{4pt}} $,
coincide
(for any not explicitly time dependent and non pathological  observable $A$).
$\gamma_\mu$ are Lagrange multipliers fixed by requiring that  the phase space averages of the
$N$ constants of motion, $\langle I_\mu\rangle_{\scaleto{\rm GGE}{4pt}}$,
equal their values evaluated at the initial conditions. For an early discussion of the GGE for a classical system
see~\cite{CuLoNePiTa18}, and for an approach based on generalised hydrodynamics 
see~\cite{Spohn,Doyon}.

Our goal here is to exhibit one such
non-trivial classical model,  the Neumann Model.
We used a mixed analytic-numerical treatment
to prove that in the thermodynamic limit, $N\to\infty$,  taken before the long-time limit, $t\gg t_{\rm st}$,
it reaches a stationary state which satisfies the extended ergodic hypothesis with a GGE measure in which the
$I_\mu$ are integrals of motion in involution  (with quartic dependencies on the phase space variables).
In so doing, we elucidate the dynamic phase diagram and we evidence condensation phenomena and
macroscopic fluctuations that should be of importance, as we explain,  in quenches of
Bose Einstein Condensates.

The Neumann Model (NM) is the simplest
non-trivial classical integrable system~\cite{Neumann}. It describes the motion of a particle on a sphere embedded in an $N$ dimensional space,
$S_{N-1}$, under fully anisotropic harmonic forces. The Hamiltonian is
\begin{equation}
H_{\rm quad} = \frac{1}{2m} \sum_\mu p_\mu^2 - \frac{1}{2} \sum_\mu \lambda_\mu s_\mu^2
\ ,
\label{eq:Hamiltonian}
\end{equation}
with $s_\mu$, $\mu=1, \dots, N$, the coordinates of the position vector,
$p_\mu$ the corresponding momentum components, $m$ the mass, and $-\lambda_\mu$ the spring constants.
The primary and secondary spherical constraints are
\begin{eqnarray}
C_1 \equiv \ \sum_\mu s_\mu^2 = N \; ,
\qquad
C_2 \equiv \ \sum_\mu s_\mu p_\mu = 0 \; .
\label{eq:constraints}
\end{eqnarray}
The equations of motion, subject to the constraints (\ref{eq:constraints}) can be derived with the
Poisson-Dirac method and read
\begin{equation}
\dot p_\mu = (\lambda_\mu - z)s_\mu \; .
\end{equation}
The ``Lagrange multiplier'' $z$ is given by
\begin{equation}
z=\frac{1}{N}\sum_{\mu}\left( p^2_{\mu}/m + \lambda_{\mu}s^2_{\mu} \right)
\; ,
\end{equation}
makes the modes interact,  and
ensures the validity of $C_1$ and $C_2$. For any initial condition satisfying these constraints, the dynamics conserve
the quadratic Hamiltonian, $H_{\rm quad}$,
as well as the $N$ Uhlenbeck integrals of motion in involution~\cite{Uhlenbeck,AvTa90,BaTa92,BaBeTa09},
\begin{equation}
I_\mu = s_\mu^2 + \frac{1}{mN} \sum_{\nu(\neq\mu)} \frac{s_\mu^2 p_\nu^2 + s_\nu^2 p_\mu^2 -2 s_\mu p_\mu s_\nu p_\nu}{\lambda_\nu-\lambda_\mu}
\; . 
\label{eq:Imu}
\end{equation}
The latter verify $\sum_\mu I_\mu = C_1$, and
$\sum_\mu \lambda_\mu I_\mu = - 2 H_{\rm pot} -   2H_{\rm kin} \, C_1/N +  1/(mN) \, C_2^2$, which equals
$-2H_{\rm quad}$ thanks to the constraints in Eq.~(\ref{eq:constraints}).

We are interested in developing a statistical description of the NM dynamics. This can make sense only in the
limit $N\to\infty$ taken before any long time limit. In this setting one can expect the
fluctuations of $z$ to be suppressed, and
\begin{equation}
z(t)\mapsto \langle z(t)\rangle_{i.c.}
\; , 
\end{equation}
where we made the time-dependencies of $z$ and its average explicit. The angular brackets represent an average over any distribution of initial conditions satisfying $\langle C_1\rangle_{i.c.} = N$ and $\langle C_2\rangle_{i.c.} = 0$. We call this variation the Soft Neumann Model (SNM). This model
has no strictly conserved quantities but $H_{\rm quad}$ and $I_\mu$, $\forall\mu$, are conserved {\it on average}.
The conditions under which the NM and SNM are equivalent will be analyzed below.

Quadratic potential energies combined with a global spherical constraint
as the  one
in Eq.~(\ref{eq:Hamiltonian}) are common
in statistical physics. Depending on the choice of the spring constants $\lambda_\mu$ one finds,  e.g.,
the celebrated spherical ferromagnet~\cite{BerlinKac,KacThompson} or the so-called $p=2$ disordered spherical model~\cite{KoThJo76,LFCDeanYoshino}. 
Problems of particles embedded in large dimensional spherical spaces and subject to random potentials
are also of this kind.
For convenience, and to make
a closer connection with the physics of disordered systems, we order the
$\lambda$'s such that $\lambda_1 < \lambda_2 < \dots< \lambda_N$ and
in the large $N$ numerical applications we take
them to be represented by a Wigner
semi-circle law on the interval $[-2J,2J]$. In this way, they can be thought of as the eigenvalues of a
two-body interaction matrix with zero mean Gaussian distributed entries
that couple the coordinates in a different basis (e.g., real spins with a global spherical constraint).
The fact that they take values within a  real interval with an edge
ensures that the total energy is bounded from below.

In most quantum quenches studies, the initial condition is taken to be the ground state of a
Hamiltonian which is suddenly modified. However,
equilibrium finite-temperature initial states~\cite{Deng11,He12,Karrasch14,Bonnes15}
are more relevant to describe, for instance, experiments in ultracold Bose gases~\cite{Eigen18}.
Along this line,
we draw the initial conditions from a proper Gibbs-Boltzmann equilibrium measure
$\rho_0=Z^{-1}_0(T') \, \exp(-\beta' H_{\rm quad}-\beta' z_{\rm eq}/2 \, (\sum_\mu s_\mu^2-N))$,
where $z_{\rm eq}$ is the equilibrium value of the Lagrange multiplier enforcing the spherical constraint
at inverse temperature
$\beta'=1/T'$ with $k_B=1$. $Z_0(T')$ is the canonical partition function,
and $H_{\rm quad}$ is given in Eq.~(\ref{eq:Hamiltonian})
with spring constants $\lambda_\mu^{(0)}$ in the interval $[-2J_0, 2J_0]$.
Depending on  $T'/J_0$ being larger or smaller than one, the initial conditions
belong to an {\it extended} phase  in which the variances of all modes are ${\mathcal O}(1)$, or  to a {\it condensed} phase
in which the averaged $N$th mode, $\langle s^2_N\rangle_{i.c.}$, scales
as ${\mathcal O}(N)$~\cite{KoThJo76}.
Two scenarii for the condensation phenomenon are possible: a mixed two pure-state
measure with the possibility of symmetry breaking induced by a vanishing pinning field,
or a Gaussian measure centered at zero with
diverging dispersion~\cite{KacThompson,Zannetti15,Crisanti-etal20}.
In the magnetic interpretation,
$T'=J_0$ is a critical point between a disordered and a magnetically ordered phase. The analogy with
Bose Einstein Condensation (BEC) was already reckoned in~\cite{KoThJo76} with $s_N$ playing the
role of the ground state density.

We drive the system out of equilibrium by performing a sudden interaction quench in which we
rescale  all spring constants, $\lambda_\mu^{(0)} \mapsto \lambda_\mu$,
with the same factor $J/J_0$ that controls the amount of energy
injected ($J/J_0<1$) or extracted ($J/J_0<1$). This procedure mimics the quenches
performed in isolated quantum systems~\cite{Polkovnikov10,Pasquale-ed16,Gogolin16}.
Right after the instantaneous quench, the initial kinetic energy of all modes
is ${\mathcal O}(1)$ and the averaged Uhlenbeck constants are
${\mathcal O}(1)$ for $T'>J_0$ while $\langle I_N\rangle_{i.c.} = {\mathcal O}(N)$ for
$T'<J_0$. Each $\langle I_\mu\rangle_{i.c.}$ is a function of $\lambda_\mu/J$ and the
adimensional parameters $T'/J_0$ and $J/J_0$ that can be easily calculated.

\begin{figure}[h!]
\centering
\includegraphics[width=7.5cm]{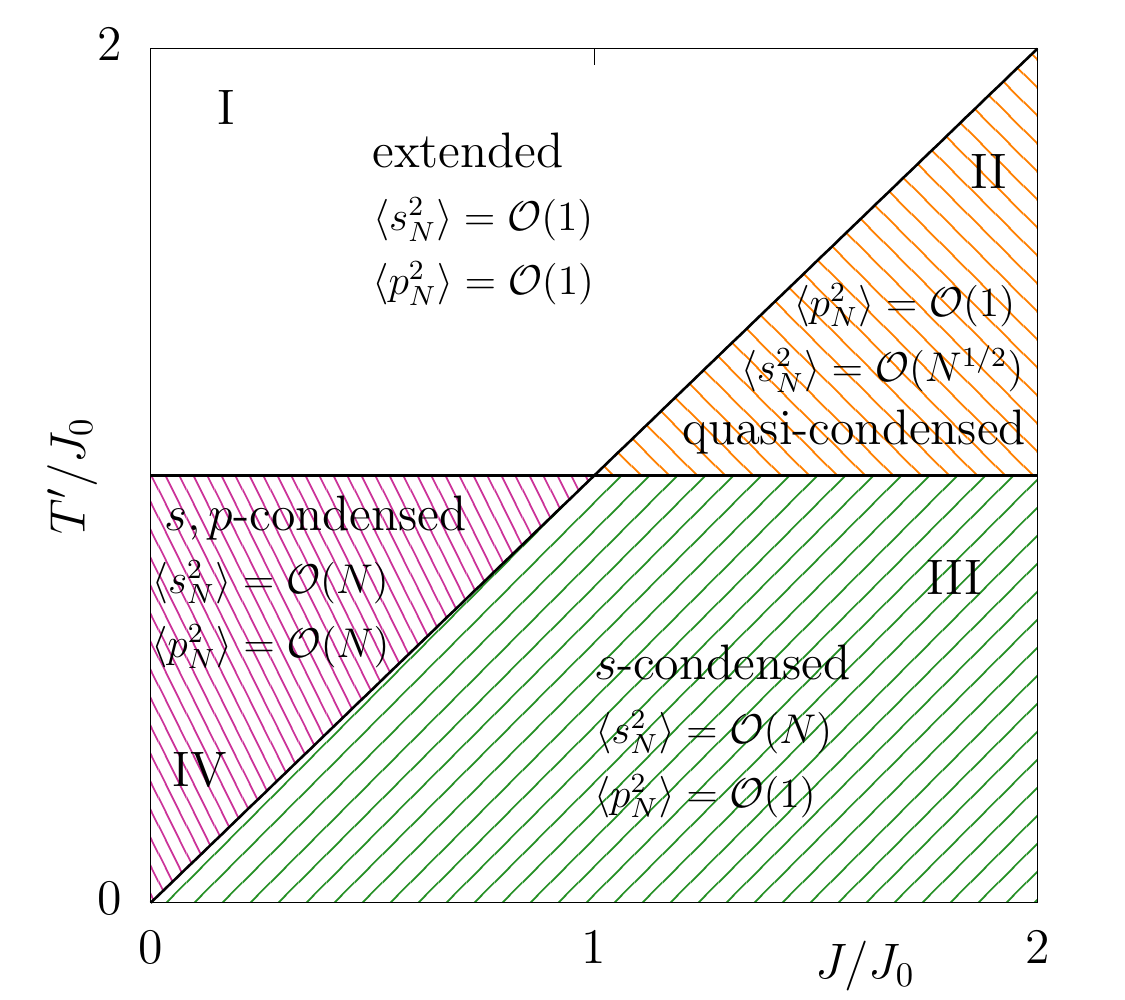}
\caption{(Color online.) The dynamic phase diagram.
$\chi_\infty = 1/T'$ to the left of the diagonal and $\chi_\infty = 1/J$
to the right of it. $q_0 \neq 0$ in III and vanishes elsewhere. The names of the
phases refer to the condensation phenomenon 
arising in III and IV, see the explanation in the text. All transition lines are continuous.
}
\label{fig:phase-diagram}
\end{figure}

Insight into the long-time dynamics  of the SNM was gained in~\cite{CuLoNePiTa18,BaCuLoNePiTa19}.
In these papers we studied the
Schwinger-Dyson equations that couple the global two-time correlation, $C$, and linear response, $R$,
averaged over the initial measure $\rho_0$ and, also,
the harmonic spring constants (quenched randomness), in the strict $N\to\infty$ limit.
This approach bears resemblance with dynamic mean theory~\cite{Aoki14}.
The (replica) method used to impose the thermal initial conditions ensures symmetry breaking for $T'<J_0$.
Four phases were identified in the $(J/J_0, \, T'/J_0)$ phase diagram (energy injection/initial condition characteristics)
as deduced from $\chi_\infty  = \lim_{t\to\infty} \int_0^t dt' \, R(t,t')$, which equals $1/J$ for $T'<J$ (II, III)
and $1/T'$ for $T'>J$ (I, IV),
and $q_0  = \lim_{t\to\infty} C(t,0)$, which takes a non-zero value for $T'<J_0$ and $T'<J$ (III), 
see Fig.~\ref{fig:phase-diagram}.  
The asymptotic value of the Lagrange multiplier is strictly larger than $\lambda_N$ for $T'>J$, whereas
it locks to $\lambda_N=2J$ for  $T'<J$ implying that the potential
on the $N$th mode flattens  and the gap of the effective Hamiltonian closes for $t\to\infty$ after $N\to\infty$.
Noteworthy, all these observables approach constant limits algebraically with superimposed oscillations~\cite{BaCuLoNePiTa19}.

In this Letter we work with  a fixed (and typical) realization of the $\lambda_\mu$.
On the one hand, we solve the coordinate dynamics for finite $N$ and, ideally, long times
with an adaptation of the semi-analytic phase-Ansatz method used in~\cite{SoCa10} to study the
O(N) field theory, and adapted in~\cite{CuLoNePiTa18} to the present case.
With this method we compute the time averages  $\overline{\langle s_\mu^2}\rangle_{i.c.}$ and $\overline{\langle p_\mu^2 }\rangle_{i.c.}$
(controlling the deviations from the ideal limit $t\to\infty$ after $N\to\infty$). On the other hand, we calculate the
GGE partition sum
\begin{equation}
Z_{\scaleto{\rm GGE}{4pt}}
= \int {\mathcal D}s \,  {\mathcal D} p\,  dz_{\scaleto{\rm GGE}{3pt}}  \; e^{-\sum_\mu \gamma_\mu I_\mu - \frac{z_{\scaleto{\rm GGE}{2.5pt}}}{2} (\sum_\mu s_\mu^2 -N)}
\; ,
\label{eq:ZGGE}
\end{equation}
with ${\mathcal D} s = \prod_\mu ds_\mu$, ${\mathcal D} p = \prod_\mu dp_\mu$
and  $z_{\scaleto{\rm GGE}{4pt}}$ the Lagrange multiplier that imposes the spherical constraint
(which in this formulation
could be reabsorbed in the definition of $\gamma_\mu$ thanks to $\sum_\mu I_\mu = C_1$).
The standard Gibbs-Boltzmann equilibrium partition
sum (relevant to describe the case $J=J_0$ and any $T'$) is recovered by
setting $\gamma_\mu=-\beta'\lambda_\mu/2$ and $z_{\scaleto{\rm GGE}{4pt}} = \beta' z_{\rm eq}$.
We evaluate the
averages $\langle s^2_\mu\rangle_{\scaleto{\rm GGE}{4pt}}$ and $\langle p^2_\mu\rangle_{\scaleto{\rm GGE}{4pt}}$ that we 
compare to the dynamic ones. We analyze the fluctuations of the constraints
$C_{1,2}$ (dynamically and with the GGE) and from their scaling we determine in which cases the SNM is equivalent to
the proper NM. 

\begin{figure}[b!]
\centering
\includegraphics[width=9cm]{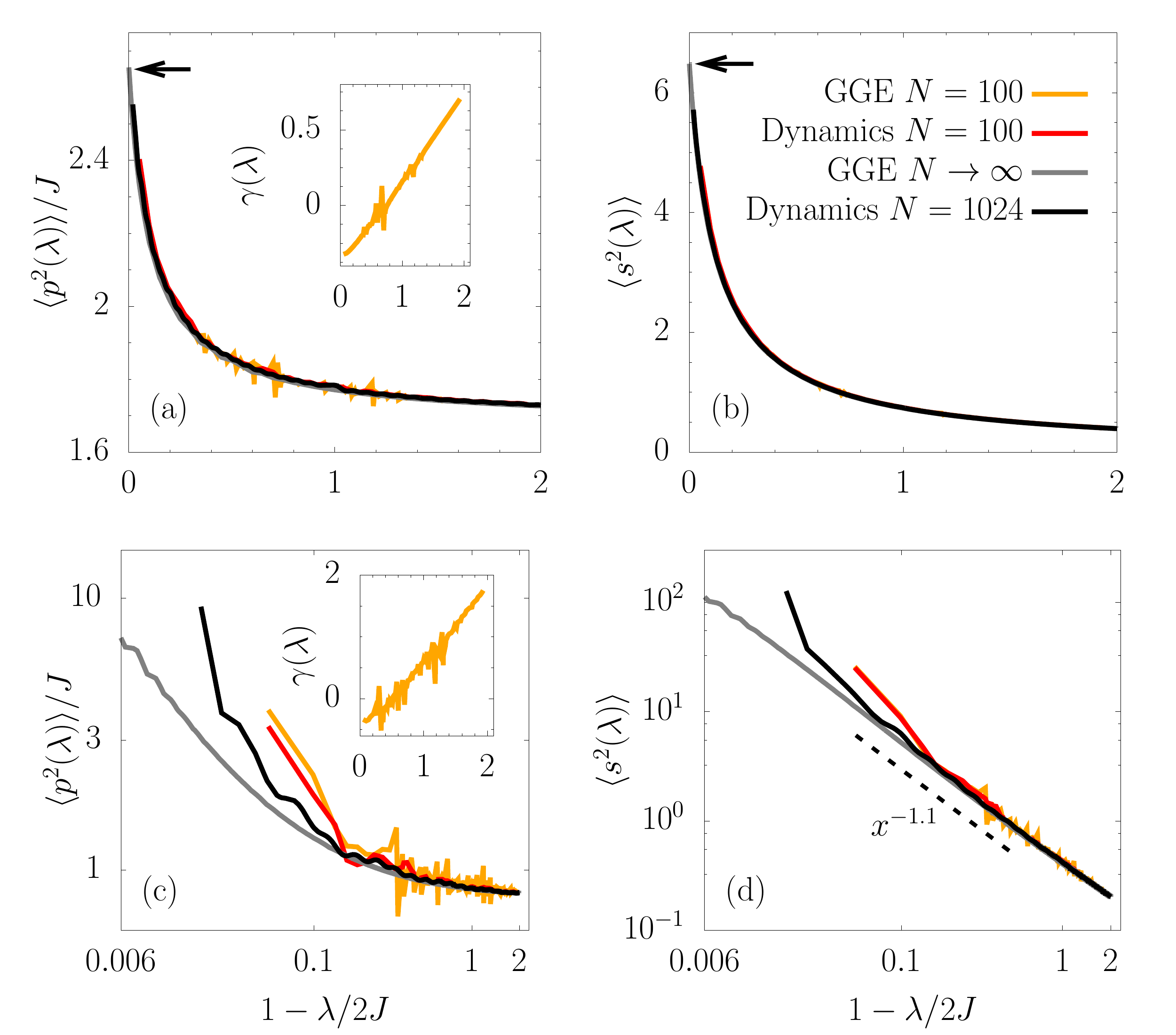}
\caption{(Color online.)
The dynamic and GGE averages of $s^2 (\lambda)$ and $p^2(\lambda)$
against $1-\lambda/2J$ in Sectors I (a) and (b), and IV (c) and (d),
of the phase diagram. In the insets the parameters $\gamma_\mu$.
The arrows in (a) and (b) indicate the finite values  of $\langle s^2 (2J)\rangle$ and $\langle p^2(2J) \rangle$
at the edge of the spectrum, contrary to their divergence in (c) and (d) (note the
double logarithmic scale). In (d) the dotted line is a guide-to-the-eye to an approximate
algebraic behavior in the bulk.
}
\label{fig:dynamics-GGE}
\end{figure}

The partition sum $Z_{\scaleto{\rm GGE}{4pt}}$  is a non-trivial object since the
$I_\mu$ are quartic functions of the phase space variables, see Eq.~(\ref{eq:Imu}). Still, we managed to calculate it by adapting
methods that are common in the treatment of disordered systems and random matrices.
Firstly, we used auxiliary variables to decouple the quartic terms.
Secondly, for $N\to\infty$,  we transformed
$\lambda_\mu$ into a continuous variable $\lambda$, all
$N^{-1} \sum_\mu A_\mu $  into $\int d\lambda \, \rho(\lambda) A(\lambda)$ for any
$A(\lambda)$, and
$\sum_{\nu (\neq \mu)} \frac{A_\nu}{\lambda_\mu - \lambda_\nu} \mapsto
\dashint d\lambda' \frac{A(\lambda')}{\lambda-\lambda'}$
with $\dashint$ the Cauchy principal value.
In some cases we separated the contribution of the $N$th mode which may be
macroscopic and  scale differently from the ones in the bulk.
Thirdly, we evaluated $Z_{\scaleto{\rm GGE}{4pt}}$ by saddle-point. Then, we showed that the harmonic {\it Ansatz}
$\langle s^2(\lambda)\rangle_{\scaleto{\rm GGE}{4pt}} = T(\lambda)/(\tilde z -\lambda)$,
$\langle p^2(\lambda)\rangle_{\scaleto{\rm GGE}{4pt}}/m =T(\lambda)$, 
solves the saddle-point equations. 
Finally, we exploit the conditions
$\langle I(\lambda)\rangle_{\scaleto{\rm GGE}{4pt}}  = \langle I(\lambda) \rangle_{i.c.} $,
with $\langle I(\lambda)\rangle_{\scaleto{\rm GGE}{4pt}} = -  \partial \ln Z_{\scaleto{\rm GGE}{4pt}} /\partial \gamma_\mu$
evaluated at the saddle point.
In the absence of  initial condition condensation, $T'>J_0$, all Uhlenbeck constants are ${\mathcal O}(1)$ and
\begin{eqnarray}
\langle I(\lambda)\rangle_{i.c.}
=
\frac{2T(\lambda)}{\tilde z -\lambda}
\;
\Big[1-\dashint d\lambda' \;  \frac{\rho(\lambda')\,T(\lambda')}{\lambda-\lambda'}\Big]
\; .
\label{eq:Uhlenbeck-bulk}
\end{eqnarray}
When the initial state is condensed,  $T'<J_0$,
Eq.~(\ref{eq:Uhlenbeck-bulk})
 applies to all $\lambda$ with the exception of $\lambda_N$, for which
\begin{equation}
 \frac{ \langle I_N\rangle_{i.c.} }
 {
2\langle s_N^2 \rangle_{{\scaleto{\rm GGE}{4pt}}}
}
=
\Big[1-\dashint d\lambda' \;
\frac{\rho(\lambda')\, T(\lambda')}{\lambda_N-\lambda'} -
\frac{\langle s_N^2 \rangle_{{\scaleto{\rm GGE}{4pt}}}}{2N}
\Big]
\label{eq:Uhlenbeck-N}
\end{equation}
plus $o(1)$ corrections. Together with the constraint $\langle C_1\rangle_{{\scaleto{\rm GGE}{4pt}}} =N$,
these are the central equations that allow us to solve the problem.
Their numerical solution
 yield the spectrum of mode temperatures,
 $T(\lambda)$, $\tilde z$ and $\langle s^2_N\rangle_{{\scaleto{\rm GGE}{4pt}}}$,
 and with them we can deduce  the expectation value of any observable.
A selected number of results are shown
in Fig.~\ref{fig:dynamics-GGE} where we compare the GGE averages to the
dynamic ones for parameters in
Sectors I and IV of the phase diagram displayed in Fig.~\ref{fig:phase-diagram}.
We collect  dynamic data for $N=100, 1024$  and GGE data for $N=100$ and $N\to\infty$.
The agreement is very good. The rather small extent of finite size effects in the bulk can also be appreciated in the
figure (the double logarithmic scale enhances the appearance of the deviations, which are actually
restricted to the neighborhood of the edge in (c) and (d)).
In the insets in (a) and (c) the spectrum of the Lagrange multipliers $\gamma_\mu$ for finite
$N$ are shown, which can be compared to the one of $T(\lambda)$.
Results of similar quality are obtained in Sectors II and III (not shown).

\begin{figure}[h!]
\begin{center}
\includegraphics[width=8.8cm]{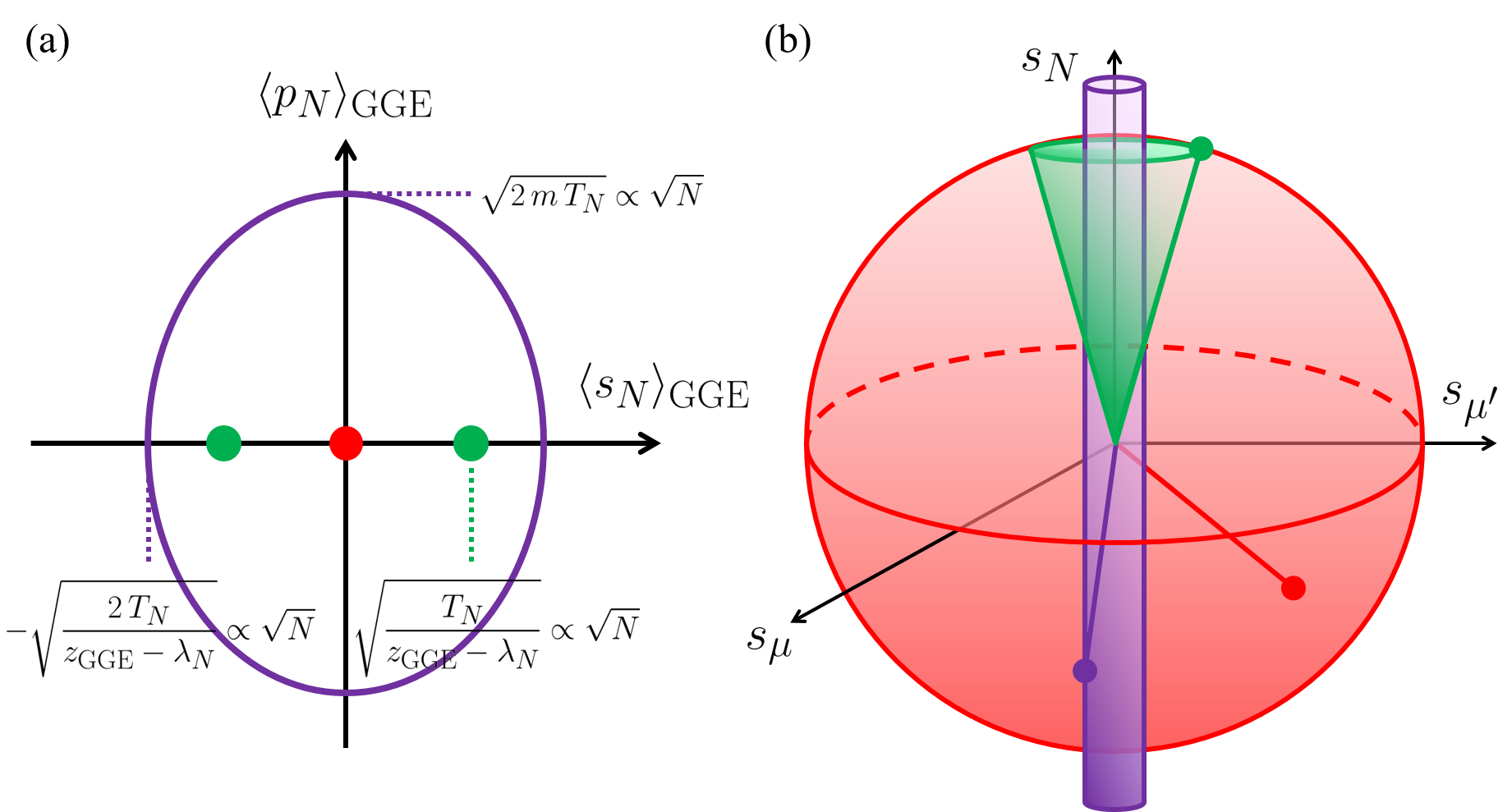}
\end{center}
\vspace{-1.25cm}
\hspace{-1.5cm}
\begin{center}
\includegraphics[width=8.8cm]{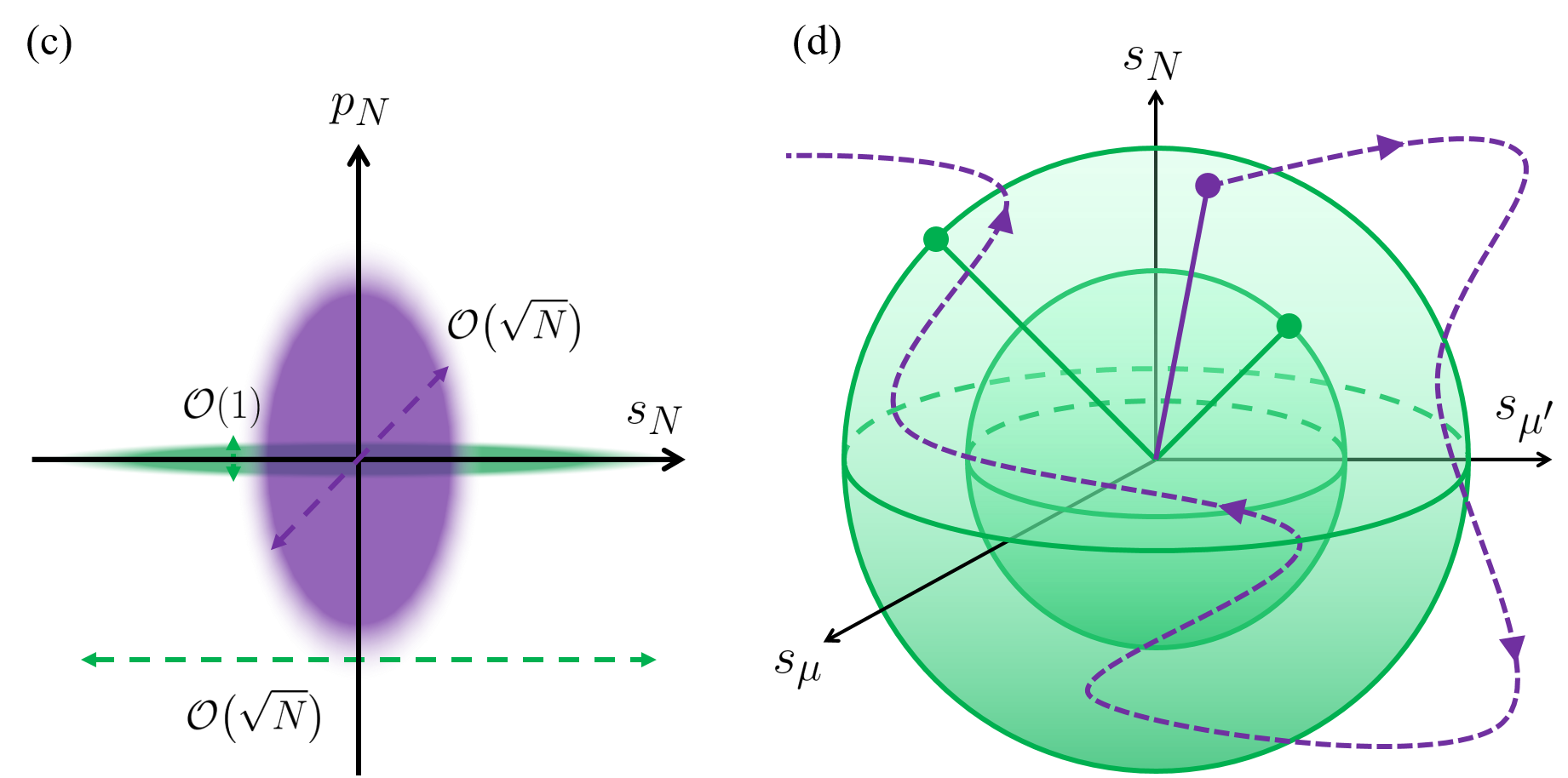}
\end{center}
\vspace{0.05cm}
\caption{
Sketches of particle trajectories in the extended (I) and quasi-condensed (II) phases  in red,
$s$-condensed (III) in green and $s,p-$condensed (IV) in violet. In
(a) and (c) we show the averaged and fluctuating $N$th mode plane, respectively, 
and in (b) and (d) the motion in the $N$-dimensional coordinate space.
The dynamics in (a) and (b) use extended ($T'>J_0$, I-II) and symmetry broken ($T'<J_0$, III-IV) initial conditions.
Panels (c) and (d) illustrate the violations of the constraints $C_i$ due to the {\it condensation of fluctuations}
for initial conditions with macroscopic fluctuations of $s_N$ (III-IV).
}
\label{fig:sketches}
\end{figure}

The dynamics in each Sector
can be rationalized according
to the scaling properties of the last mode and the fluctuations of the
constraints
\begin{eqnarray}
\Delta C_i \equiv
\left\langle (C_i -  \left\langle C_i \right\rangle)^2 \right\rangle
\qquad i=1,2
\; ,
\end{eqnarray}
which can be studied both dynamically and with the GGE.
When the scaling of these fluctuations is ${\mathcal O}(N^2)$ the SNM is not equivalent to the NM.

In Sector I,
$\langle s^2_{\mu}\rangle_{\scaleto{\rm GGE}{4pt}}$ and $\langle p^2_{\mu}\rangle_{\scaleto{\rm GGE}{4pt}}$ are $\mathcal{O}(1)$ for all $\mu$, including $\mu=N$. In a sense, this is the simplest possible
generalization of the Boltzmann equilibrium {\it extended} phase.
In Sector II, we have numerical evidence for $\langle s^2_{N} \rangle_{\scaleto{\rm GGE}{4pt}}$
scaling as $N^{1/2}$, while $\langle s^2_{\mu\neq N}\rangle_{\scaleto{\rm GGE}{4pt}}$ and $\langle p^2_{\mu}\rangle_{\scaleto{\rm GGE}{4pt}}$ should
be $o(N^{1/2})$.
This is a {\it quasi-condensed} phase in which the weight of the last mode is large but not extensive. Since there is no condensation, the energy conserving dynamics in the extended and quasi-condensed phases explore the full sphere in the course of time as sketched in Fig.~\ref{fig:sketches}(a),(b) with a
red dot and the red sphere, respectively. Moreover, $\Delta C_i = o(N^2)$ and the NM and SNM models are equivalent.

As explained above,  the initial conditions drawn from the Boltzmann measure of the SNM at $T'<J_0$ can be of two kinds:
(i) $s_N(t=0) \propto \pm \sqrt{N}$ with negligible fluctuations, or (ii) $s_N(t=0) $ Gaussian distributed, centered
at zero with $\sqrt{N}$ fluctuations~\cite{KacThompson,Crisanti-etal20}.
In both cases $\langle s_N^2\rangle_{i.c.} \propto N$, but the ensuing dynamics are different and have to
be discussed separately.

In case (i), Sector III is a properly $s$-{\it condensed}
phase with $\langle s^2_{N}\rangle_{\scaleto{\rm GGE}{4pt}}$ scaling as $N$,
while $\langle s^2_{\mu\neq N}\rangle_{\scaleto{\rm GGE}{4pt}} = o(N)$
and $\langle p^2_{\mu}\rangle_{\scaleto{\rm GGE}{4pt}}=\mathcal{O}(1)$.
The system precesses around one of the two states with
 $|s_N| = {\mathcal O}(N^{1/2})$, the one selected by the symmetry broken initial conditions,
 and comparably negligible projection on all other directions, see
 the symmetrically placed green dots and green trajectory in Fig.~\ref{fig:sketches}(a),(b), respectively.
 The constraints $C_1$ and $C_2$ are strictly satisfied up to sub-extensive corrections
 and the NM and SNM models are equivalent.
Remarkably,  in Sector IV both $\langle s^2_{N}\rangle_{\scaleto{\rm GGE}{4pt}}$
and $\langle p^2_{N}\rangle_{\scaleto{\rm GGE}{4pt}}$ scale as $N$, and
the $N$th mode captures ${\mathcal O}(N)$ kinetic energy.  We call this Sector an
$s,p$-{\it condensed} phase.
The last mode is in a superposition of states associated to each initial condition. At any instant  $t$,
the configurations are distributed on an ellipse in the plane
$(s_N,p_N)$ with axes  ${\mathcal O}(N^{1/2})$, as in the closed motion of a harmonic
oscillator, see the violet ellipse and
cylinder in Fig.~\ref{fig:sketches}(a),(b), respectively. The average over trajectories implies, in particular, that the limit correlation $q_0$ vanishes.
The constraints $C_{1,2}$ are only verified on average over the initial conditions and the 
SNM and NM models are not equivalent. 
We note that $\Delta C_{1,2}$ are averages of a quartic functions of the phase variables; had we evaluated
only quadratic functions of $s_N$ we would have not noticed the inequivalence between the two models.
Quite surprisingly, the averaged dynamics cannot be boiled down to the ones
of a typical trajectory with its own $z(t)$.

In case (ii), the initial conditions imply $\Delta C_1 = {\mathcal O}(N^2)$
at all times due to the large fluctuations of the last mode. One can show that, in
Sector  III, $\Delta C_2 = o(N^2)$ at all times. 
In this situation, due to the large fluctuations in $C_1$, zero-mean initial 
conditions are appropriate for the soft model but not for the strictly spherical one.
In practice, in the SNM we average over spherical trajectories with 
different radius determined by the initial condition.
In Sector IV, due to the condensation of $p_N$, the dynamics do not preserve the scaling properties of $C_2$ either.
In other words, the fluctuations of the secondary constraint, which vanish in the initial condition,
get macroscopically amplified by the dynamics. In conclusion, we average over trajectories that no longer move on the
sphere. In this Sector, the fluctuations of all the quantities that are conserved {\it on average},
$H_{\rm quad}$, $C_{1,2}$ and $I_N$, condense, which implies  that the dynamics do not conserve the quadratic energy,
are not restricted to a sphere and are not strictly integrable.
The behaviours in Sectors III and IV 
are represented in Fig.~\ref{fig:sketches}(c),(d), with the same
colour code as the one we used before.

Contrary to the  quantum mechanical subtleties~\cite{Caux10,Yuzbashyan11},
the notion of classical integrability is clear~\cite{Dunajski12,Arnold78,Khinchin}. 
The dynamics should be ergodic on the portion of
phase space compatible with the constants of motion~\cite{Yuzbashyan16}.
Still, the fact that a canonical
GGE could describe the time-averages of generic observables in a classical interacting integrable system is not
obvious. We modified the celebrated Neumann model by imposing the spherical constraint on average
over the initial conditions and we were then able to solve it  in the thermodynamic limit. We
thus provided an explicit example in which identities between temporal and statistical averages, 
for all kinds of thermal initial conditions (on average) and observables not correlated with the
constants of motion and post-quench parameters,
can be demonstrated. Importantly enough, for condensed
initial states,  $\langle s^2_N\rangle_{i.c.}$ and $\langle I_N\rangle_{i.c.}$
are macroscopic and stay so after the quench. 
In these cases, we distinguished symmetry
broken initial conditions and  symmetric ones with zero mean and condensed fluctuations.
Quadratic observables are insensitive to the changes that the latter induce
but quartic ones are not. 
For symmetry broken initial conditions, the SNM behaves just as the NM in the phase
in which only the coordinate is condensed but it loses its equivalence with the NM in the phase in
which not only the coordinate but also the momentum condenses. 
For  initial states with macroscopic fluctuations,
integrability is valid only on average over initial conditions.
Energy conservation is violated in the
condensed Sectors of the phase diagram and the SNM and NM models are not equivalent.
Interestingly enough, given the similarity between the phase transitions and condensation in this model and in
BEC~\cite{KoThJo76,Crisanti-etal20} we may expect similar phenomena in quenches of thermal
initial states of the latter.

\noindent
{\it Acknowledgments.}
We thank J-B Zuber for very helpful discussions.

\bibliographystyle{apsrev4-1}

\begin{thebibliography}{99}
\bibitem{footnote}
The time $t_{\rm st}$ is the time-scale needed to reach stationarity and
it will typically be much longer than a microscopic time-scale $t_0$.
\end{thebibliography}

\end{document}